\documentclass[conference]{IEEEtran}
\IEEEoverridecommandlockouts
\usepackage{cite}
\usepackage{enumitem}
\usepackage{threeparttable} 
\usepackage{array}
\usepackage{subfigure}
\usepackage{tablefootnote}
\usepackage{booktabs}
\usepackage{makecell}
\usepackage{stfloats}
\usepackage{amsmath,amssymb,amsfonts}
\usepackage{graphicx}
\usepackage{textcomp}
\usepackage{xcolor}
\usepackage{pifont}
\usepackage{algorithm}  
\usepackage[normalem]{ulem}  
\usepackage{algpseudocode}  
\usepackage{amsmath}  
\usepackage{multirow}
\usepackage{fix-cm}

\usepackage{url}

\usepackage{geometry}

\begin{document}

\title{\fontsize{17.9}{22}\selectfont \bf HLSTester: Efficient Testing of Behavioral Discrepancies with LLMs for High-Level Synthesis\vspace{-0.35cm}}

\author{
\IEEEauthorblockN{Kangwei Xu\textsuperscript{1}, Bing Li\textsuperscript{2}, Grace Li Zhang\textsuperscript{3}, Ulf Schlichtmann\textsuperscript{1}}
\IEEEauthorblockA{\textsuperscript{1}\textit{Chair of Electronic Design Automation, Technical University of Munich (TUM)}, Munich, Germany \\
\textsuperscript{2}\textit{Research Group of Digital Integrated Systems, University of Siegen}, Siegen, Germany \\
\textsuperscript{3}\textit{Hardware for Artificial Intelligence Group, Technical University of Darmstadt}, Darmstadt, Germany \\
Email: kangwei.xu@tum.de, bing.li@uni-siegen.de, grace.zhang@tu-darmstadt.de, ulf.schlichtmann@tum.de}
\vspace{-0.95cm}
}

\maketitle
\thispagestyle{empty}

\begin{abstract}
In high-level synthesis (HLS), C/C++ programs with synthesis directives are used to generate circuits for FPGA implementations. However, hardware-specific and platform-dependent characteristics in these implementations can introduce behavioral discrepancies between the original C/C++ programs and the circuits after high-level synthesis. 
Existing methods for testing behavioral discrepancies in HLS are still immature, and the testing workflow requires significant human efforts. 
%Existing methods for testing behavioral discrepancies in HLS still require significant human efforts. 
To address this challenge, we propose \textbf{HLSTester}, a large language model (LLM) aided testing framework that efficiently detects behavioral discrepancies in HLS. To mitigate hallucinations in LLMs and enhance prompt quality, existing C/C++ testbenches are used to guide the LLM to generate HLS-compatible versions, effectively eliminating certain traditional C/C++ constructs that are incompatible with HLS tools. Key variables are pinpointed through a backward slicing technique in both C/C++ and HLS programs to monitor their runtime spectra, enabling an in-depth analysis of the discrepancy symptoms. Then, a testing input generation mechanism is introduced to integrate dynamic mutation with insights from an LLM-based progressive reasoning chain. In addition, repetitive hardware testing is skipped by a redundancy-aware technique for the generated test inputs. Experimental results demonstrate that the proposed LLM-aided testing framework significantly accelerates the testing workflow while achieving higher testbench simulation pass rates compared with the traditional method and the direct use of LLMs on the same HLS programs.

\end{abstract}

\section{Introduction}
As the demand for hardware performance in AI computing continues to grow, an increasing number of software applications originally designed for CPUs are being ported to specialized hardware accelerators like FPGAs \cite{b1,b2,b3}. High-Level Synthesis (HLS) tools can take general-purpose programming languages such as C/C++ and generate Hardware Description Language (HDL) designs, thereby synthesizing corresponding circuits for FPGA deployment \cite{b4, b4.1, b4.2}.

Due to limitations of HLS tools, only a subset of C/C++ programs is supported, referred to as HLS programs \cite{b5}. When these programs are synthesized into circuits by the HLS tool, they may exhibit behavioral discrepancies compared with the original C/C++ programs running on CPUs. Such discrepancies include overflow from customized bit width, out-of-bounds memory accesses caused by static allocation, data dependencies due to pipelining, etc. For example, default data types on CPUs are typically represented as \textit{int} or \textit{float}, whereas in FPGA designs, the bit width of integers and floats are often customized, which could lead to overflows on FPGAs. Additionally, FPGAs enable parallel computing through pipeline execution by incorporating \textit{hardware directives} such as \texttt{\#pragma} HLS pipeline, which may produce results that differ from those of sequential CPU execution, particularly in cases where data dependencies or feedback paths exist among functions. 

Given these potential divergences, systematic testing is essential in the HLS workflow, as programs that execute correctly on CPUs may still fail or produce incorrect results when deployed on FPGAs \cite{b7}. To test behavioral discrepancies in HLS, developers typically need to monitor runtime spectra by constructing testbenches and instrumenting key variables. The term \textit{spectra} refers to dynamic traces that capture variable values, memory access patterns, stack sizes, FIFO queue sizes, and other relevant runtime metrics. Any discrepancies in these spectra between C/C++ programs on CPUs and synthesized circuits on FPGAs can reveal issues in HLS that need to be addressed \cite{b7.8}. 
However, this testing process requires developers to have expertise in both hardware and software, with significant efforts in the overall testing workflow. 
In addition, developers generally employ fuzz testing \cite{b8}, a common test input generation scheme that has been widely applied in software testing where each test run completes in just a few milliseconds\cite{b8.1}. However, in hardware testing, particularly in FPGA testing where clock-cycle-level simulation is required due to data dependencies, the testing latency (i.e., the time required to perform a test operation) often ranges from several minutes to several hours, rather than milliseconds\cite{b8.2}. Moreover, test inputs generated by fuzz testing do not take hardware platform characteristics into account, making it ineffective in testing behavioral discrepancies in HLS.

To overcome these limitations, more advanced testing methodologies are emerging. Large language models (LLMs) have demonstrated great potential in automated software and hardware tasks, thereby supporting the entire process from initial design to testing, verification, and physical implementation \cite{b11,b11.4,b11.6,b11.7,b11.8,b11.9,b11.10}. Recent research has shown that LLMs can be employed to test errors in C/C++ programs and Hardware Description Language (HDL) designs. CityWalk \cite{b12} leverages project-dependency analysis and insights from documentation to address complex features, such as templates and virtual functions, thus producing high-quality testbenches. CorrectBench \cite{b13} employs LLMs to implement a testing platform with self-correction, and an automatic validation framework is proposed to evaluate the performance of the generated testbenches. %VerilogReader \cite{b14} employs LLMs as readers to understand Verilog code, aiming to generate code convergence tests. In addition, a testing interpreter is introduced to enrich prompts and thus enhance LLM’s understanding of the testing intent.

Previous methods above use LLMs to test simple errors in common code types, such as individual C/C++ or Verilog programs that LLMs are extensively trained on and thus highly familiar with. However, these methods do not provide a testing workflow to detect behavioral discrepancies in interdisciplinary fields such as HLS, which necessitates the co‑exploration of hardware and software \cite{b11}. 
In this paper, we introduce \textbf{HLSTester}, an LLM‑aided HLS testing framework that employs an iterative testing workflow to automatically detect behavioral discrepancies in the HLS design. The key contributions of this paper are summarized as follows.

\begin{table}[t]
\vspace{0.1cm}
\centering	
  \refstepcounter{table}%
  {TABLE I: \MakeUppercase{Typical discrepancy sources when porting C/C++ programs from CPU to FPGA via High-Level Synthesis.\par}}
\vspace{0.25cm}
 \includegraphics[width=1\linewidth]{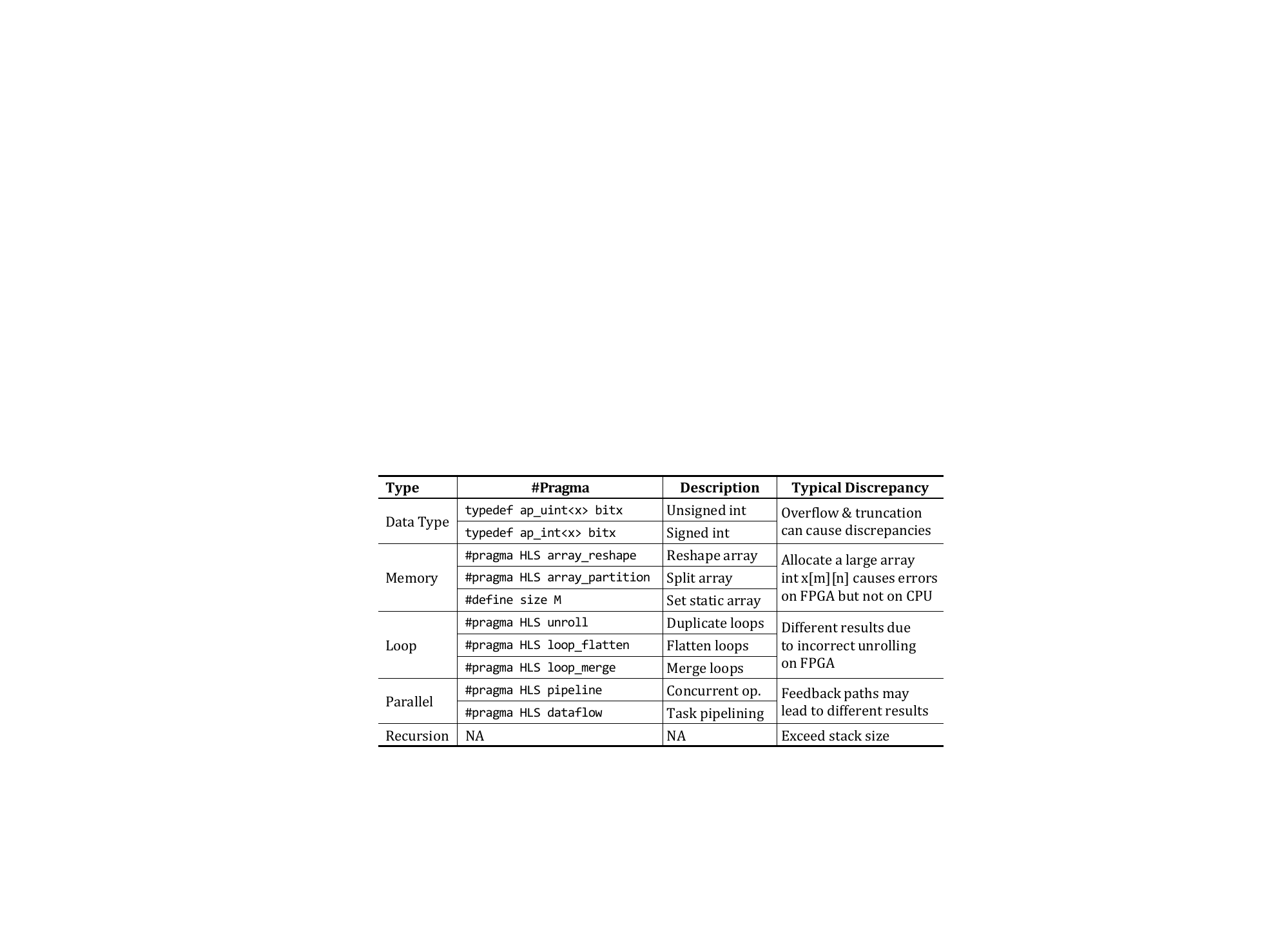}
	\label{fig:type}
	\vspace{-0.61cm}
\end{table}

\begin{itemize}
\item 
We propose an efficient testing framework for HLS that covers code analysis, testbench modification, and runtime profiling for behavioral discrepancy detection.

\item 
By leveraging the adaptive learning capability of LLMs, existing C/C++ testbenches are used to guide the LLM to generate HLS-compatible versions, improving the pass rate of the resulting testbenches by an average of 20.67\%.

\item 
To enable an in-depth analysis of the discrepancy, a backward slicing technique, combined with the semantic understanding capabilities of LLMs, is employed to instrument the HLS program. Monitoring scripts are also developed to record the runtime spectra of key variables, which could also provide a basis for test input mutation.

\item 
A test input generation mechanism is proposed that leverages spectra feedback to guide dynamic mutation, along with insights from an LLM-based progressive reasoning chain. The LLM-guided test input generation achieves an average of 1.81× speedup in detecting all behavioral discrepancies, while the dynamic mutation further increases the speedup to an average of 2.28x.

\item 
To avoid redundant hardware simulations during HLS testing, a redundancy-aware technique is proposed to identify and skip repetitive test inputs, which contributes an additional 15.73\% acceleration in the testing workflow.

\end{itemize}

The rest of this paper is organized as follows. Section~\ref{sec:second} provides the background and motivation for this work. Section~\ref{sec:third} details the proposed framework. The experimental results are provided in Section~\ref{sec:fourth}. Section~\ref{sec:fifth} concludes the paper.

\vspace{-0cm}
\section{Background and Motivation}\label{sec:second}
Since some traditional C/C++ constructs are not supported by HLS tools or lead to inefficient synthesized circuits, significant code refactoring and the incorporation of specific hardware directives are necessary to adapt them for synthesis\cite{b5}. Testing behavioral discrepancies between the original C/C++ programs and the circuits after high-level synthesis is challenging due to assumptions made during synthesis.
Table~\ref{fig:type} presents typical examples of how directives cause these discrepancies \cite{b6}.

$\circ$~\textit{Data Type:} HLS supports arbitrary bit width for variables, leading to a reduction in FPGA resource usage. For example, analysis of the variable ‘m' in a standard C++ program exhibits a minimum value of 0 and a maximum of 481, thereby necessitating only 9 bits instead of the 32-bit \textit{int} type. In such cases, the directive \texttt{\#typedef ap\_uint$<$9$>$} is used to declare a 9-bit precision integer. With custom bit width, FPGAs become more susceptible to overflow, leading to incorrect results.

\begin{figure}[t]
\vspace{-0.16cm}
\centering	\includegraphics[width=1.02\linewidth]{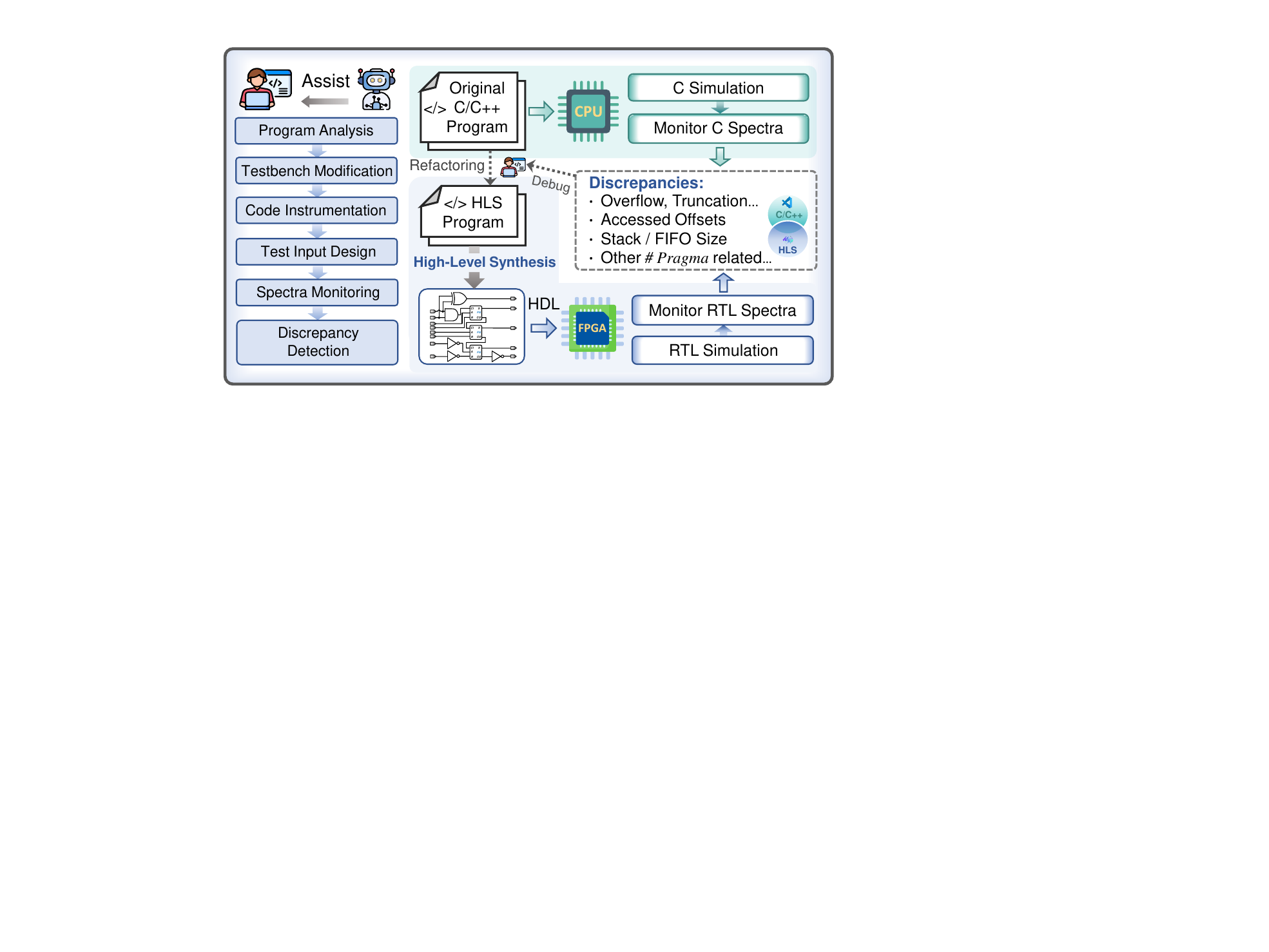}
	\vspace{-0.69cm}
	\caption{~Traditional workflow for testing behavioral discrepancies in HLS.}
	\label{fig:trad}
	\vspace{-0.67cm}
\end{figure}

$\circ$~\textit{Memory Management:} HLS tools do not support dynamic arrays because FPGAs are unable to manage unbounded data structures. As a result, functions such as \texttt{malloc()} should be replaced with static arrays. Accessing an unexpected address may cause the FPGA to return an incorrect output.

$\circ$~\textit{Loop and Parallelization:} Hardware directives such as \texttt{\#pragma HLS unroll} create multiple instances of a loop body in the generated RTL design, enabling parallel execution of loop iterations. 
In addition, the directive \texttt{\#pragma HLS dataflow} enables task-level pipelining, allowing functions to overlap their operations \cite{b14.4}. However, such parallel execution can produce results that differ from sequential CPU execution, particularly when misaligned data accesses occur, or feedback paths exist between different variables or functions.

$\circ$~\textit{Recursion:} HLS tools do not support dynamic stack allocation for execution states, so recursion needs to be converted to iteration using a finite-size stack \cite{b14.5}. When recursion depth exceeds the estimated size, FPGA compilation failures occur.

An overview of the traditional workflow for testing behavioral discrepancies in HLS is shown in Fig.~\ref{fig:trad}. Typically, developers begin with a program analysis, followed by modifications to the testbench to eliminate certain traditional C/C++ constructs that are incompatible with HLS tools. Then, both the original C/C++ programs and their HLS counterparts are instrumented, and a set of relevant test inputs is designed to be applied identically in both software and hardware simulations. At runtime, spectra of key variables are recorded and compared between the original C/C++ program and the circuits after high-level synthesis, and any mismatches in these spectra indicate behavioral discrepancies. On the right side of Fig.~\ref{fig:trad}, such discrepancies may arise from overflows, truncations, and issues introduced by other hardware directives, which are then used to debug the refactored HLS program.

\begin{figure}[t]
\vspace{-0.07cm}
\centering	
\includegraphics[width=0.82\linewidth]{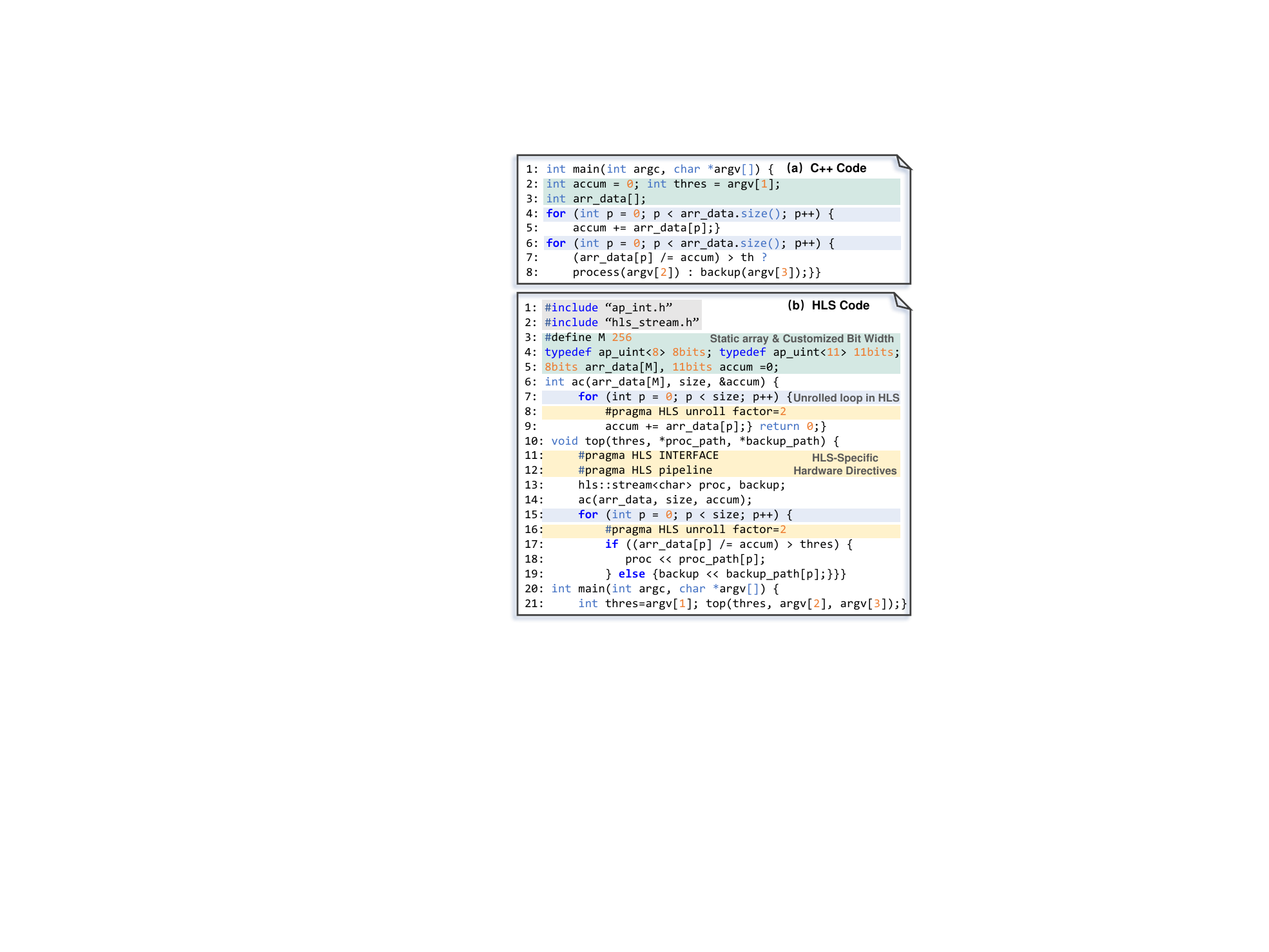}
\vspace{-0.34cm}
\caption{~Discrepancies between (a) the original C++ code \& (b) the HLS code.}
\label{fig:trans}
\vspace{-0.6cm}
\end{figure}

To illustrate the need for testing behavioral discrepancies in HLS environments, Fig.~\ref{fig:trans}(a) presents an edge processing task. In this task, the original C++ program first computes the cumulative sum of an image gradient vector and then compares the percentage of each gradient against a threshold to decide whether to process pixels. In Fig.~\ref{fig:trans} (b), to address the performance bottleneck on the CPU and fully exploit FPGA parallelism, the developer refactored the C++ code, which introduced custom bit width, static arrays, and various hardware directives. However, these hardware-specific refactorings may induce platform-dependent behavioral discrepancies. For example, custom bit width could result in overflow or truncation issues; replacing dynamic memory allocation with static arrays may lead to out-of-bound accesses; and the pipelining may produce results that differ from sequential CPU execution.

To generate these test inputs in the HLS workflow, several methods are typically employed: (1) random generation using data generators \cite{b15}; (2) handcrafted inputs by an HLS expert; or (3) enumeration through scripts \cite{b16,b16.0,b16.01}. Typically, the test input generation begins with a set of seed inputs, selecting one seed randomly and producing new inputs by altering a few bits. 
However, in FPGA testing, a single hardware synthesis and testing simulation may take several minutes or even hours. Moreover, traditional mutation techniques in software testing overlook hardware-specific characteristics such as custom bit width, static allocation, and pipelining via hardware directives, resulting in ineffective behavioral discrepancy testing in HLS.

Contrary to previous work, this paper introduces \textbf{HLSTester}, a framework that leverages the in-context learning and comprehension capabilities of LLMs to assist key testing points. Integrating testing scripts with feedback from HLS tools accelerates discrepancy testing and leads an efficient testing workflow.

\vspace{-0.1cm}
\section{LLM-Aided Behavioral Discrepancy Testing}\label{sec:third}

\begin{figure}[t]
\vspace{-0.07cm}
\centering
	\includegraphics[width=1\linewidth]{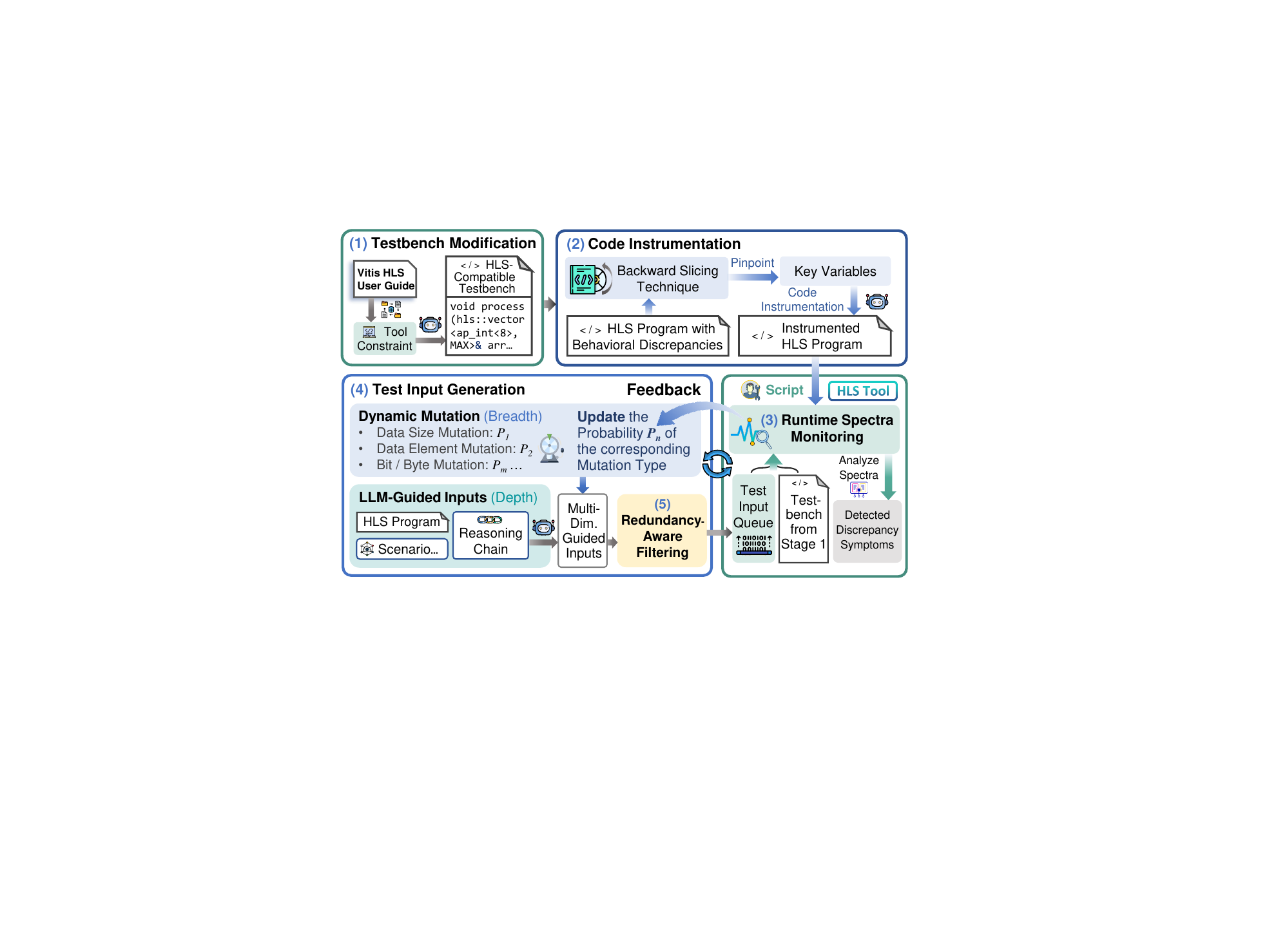}
	\vspace{-0.7cm}
	\caption{Testing Workflow of the proposed HLSTester.}
	\label{fig:wf}
	\vspace{-0.7cm}
\end{figure}

The proposed testing framework is composed of five stages: (1) testbench modification for HLS, (2) code instrumentation, (3) runtime spectra monitoring, (4) test input generation, and (5) redundancy-aware filtering. As illustrated in Fig.~\ref{fig:wf}, the framework first constructs a testbench compatible with the HLS tool by eliminating incompatible C/C++ constructs. Next, a backward slicing technique is employed to pinpoint key variables, which are then instrumented to enable monitoring of their runtime spectra. The collected spectra are fed back into the test input generation stage to guide the dynamic mutation of the test inputs, forming a cyclic feedback system. Finally, a redundancy-aware technique for the generated test inputs is applied to skip duplicate hardware simulations. Each stage will be described in detail in the following subsections. In the following description, we use the terms below in the HLS flow:

\begin{itemize}
  \item \textbf{Original C/C++ program}: The original software implementation written in C/C++ designed for CPU execution.
  \item \textbf{HLS program}: The HLS-compatible program converted from the original C/C++ program with directives.
  \item \textbf{Original testbench}: The original C/C++ testbench developed to test the original C/C++ program.
  \item \textbf{HLS-compatible testbench}: A testbench modified from the original C/C++ testbench and able to be compiled by HLS tools to test the generated RTL design.
\end{itemize}

\vspace{-0.2cm}
\subsection{Testbench Modification for HLS}
\vspace{-0.05cm}
%Certain C/C++ constructs in C/C++ testbenches, such as traditional C/C++ header files, definitions for dynamic arrays, and other unsupported features, are incompatible with HLS tools. However, despite the strict constraints imposed by HLS tools, both types of testbenches exhibit notable similarities in their intended functionality and basic structure. To enable the modification of a C/C++ testbench into an HLS-compatible version, three key components are prepared: (1) the original C/C++ program, (2) its corresponding C/C++ testbench, and (3) a refactored version of the program, referred to as the HLS program under test, which complies with the constraints of the HLS tool. At this stage, given the adaptive learning capabilities of LLMs, existing C/C++ testbenches are used as references to modify them into versions compatible with HLS tools. We formalize the testbench modification for HLS as follows:

To test the discrepancies between a given C/C++ program and the HLS program converted from it, a testbench is needed. However, the original C/C++ testbench cannot be used directly, as it may contain C/C++ constructs that are incompatible with HLS tools and thus cannot be compiled to test the target design, e.g., traditional C/C++ header files, definitions for dynamic arrays, and other unsupported features. To deal with this problem, we use an LLM to modify the original C/C++ testbench to generate the HLS-compatible testbench.
% At this stage, given the adaptive learning capabilities of LLMs, existing C/C++ testbenches are used as references to modify them into versions compatible with HLS tools. 
%The procedure of this testbench modification for HLS can be described as follows:
%
%\vspace{-0.17cm}
%\begin{equation}
%\label{eq:LLM_transform}
%F_{\mathrm{LLM}} : T_{\mathrm{C/C++}}\;\;\to\;\;T_{\mathrm{HLS}}, 
%\quad \text{s.t.} \quad \mathrm{Constraints}\_{\mathrm{HLS},}
%\vspace{-0.17cm}
%\end{equation}
%where $T_{\mathrm{C/C++}}$ is the set of traditional C/C++ testbenches, $T_{\mathrm{HLS}}$ is the set of HLS-compatible testbenches, $\mathrm{Constraints}\_{\mathrm{HLS}}$ are the constraints imposed by HLS tools. %To achieve automated generation of an HLS testbench, 
%
The LLM focuses on the following aspects during the analysis and modification process to generate the HLS-compatible testbench:

(1) Code Semantic Analysis: The LLM performs a semantic analysis of the C++ testbench to understand its core logic, functionality, and structure. This step ensures that the LLM fully captures the key functionalities and operations in modifying a testbench that is compatible with the HLS tool.

(2) Usage limitations: To meet HLS tool requirements, the LLM adjusts the header files and data types in the original C/C++ testbench to match the definitions provided in the HLS User Guide. For instance, standard data transfers in C++ need to be changed to stream-based data transfers by including the \texttt{$<$hls\_stream.h$>$} header file and using the \texttt{hls::stream} to emulate hardware stream processing structures. Data types should also be adapted for the HLS program, such as replacing the \texttt{int} type in the C++ testbench with the \texttt{ap\_int} type.

\begin{figure}[t]
\vspace{-0cm}
\centering
	\includegraphics[width=1\linewidth]{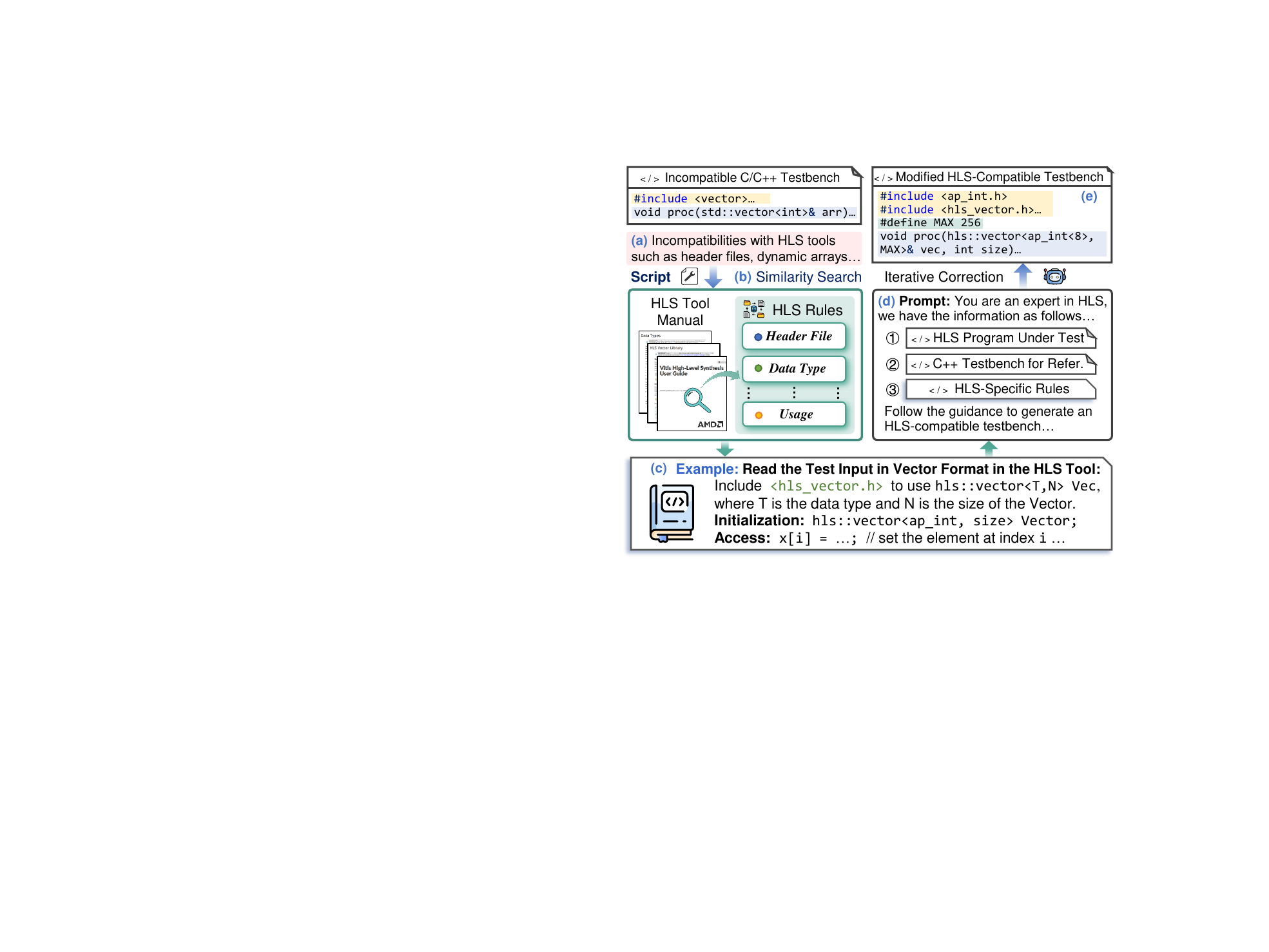}
	\vspace{-0.7cm}
	\caption{LLM-aided testbench modification for HLS.}
	\label{fig:tbg}
	\vspace{-0.65cm}
\end{figure}

To enhance LLM prompts, a Retrieval-Augmented Generation (RAG) technique is employed by incorporating external guidance via a retriever. This process involves three stages:

First, an external rule library is created containing HLS rule templates. Each template includes potential error logs that may be encountered during synthesis, a summary of the associated HLS rule, and usage examples extracted from the HLS manual.

Second, the existing C/C++ testbench is compiled with the HLS tool, and its error logs are used as queries to retrieve templates from the rule library. These rule templates embed potential error logs that not only facilitate matching with the compiler-generated error logs but also incorporate guidelines for correcting these errors. A sentence transformer \cite{b17} is used to convert the error log and each rule template into embedding vectors. The retrieval process can be formalized as follows:
\vspace{-0.15cm}
\begin{equation}\label{eq:define}
E = \mathrm{embed}(\text{error\_log}), \quad
R_i = \mathrm{embed}(\text{rule\_template}_i),
\vspace{-0.15cm}
\end{equation}
where \(E\) and \(R_i\) represent the embedding vectors of the compiler-generated error log and the \(i\)th rule template. Given an error log embedding $E$ and a rule template embedding $R$, their similarity is computed via cosine similarity, and the rule template with the highest similarity is selected as follows:
\vspace{-0.1cm}
\begin{equation}\label{eq:cosine_selection}
\mathrm{sim}(E,R) = \frac{E \cdot R}{\|E\|\,\|R\|}~, \quad
R^{*} = \underset{R_i \in \mathcal{R}}{\operatorname{arg\,max}}\, \mathrm{sim}(E,R_i),
\vspace{-0.1cm}
\end{equation}
where $\mathcal{R}$ denotes the set of all rule templates in the library, $R^{*}$ is the rule template from $\mathcal{R}$ that has the highest cosine similarity with the compiler-generated error log embedding \(E\).

Third, once the template with the highest similarity is retrieved, it is incorporated into the LLM prompt to guide the modification of the testbench into an HLS-compatible version.

An example of applying RAG to modify a C/C++ testbench into an HLS-compatible version is shown in Fig.~\ref{fig:tbg}. Initially, when the HLS compiler detects incompatible errors (a), such as those related to the use of \texttt{$<$vector$>$} and dynamic arrays, it generates error logs. These error logs are then used to retrieve the most appropriate rule template from the external library via a sentence transformer (b). Once the rule template with the highest similarity is retrieved (c), for example, one that recommends replacing \texttt{$<$vector$>$} with \texttt{$<$hls\_vector.h$>$}, it is incorporated into the prompt provided to the LLM (d). 
The LLM then follows this guideline to generate a new testbench (e). If the HLS compiler still detects an error, the process iteratively repairs the testbench until a correct version that conforms to HLS constraints is obtained.

\begin{figure}[t]
\vspace{-0.03cm}
\centering
	\includegraphics[width=1\linewidth]{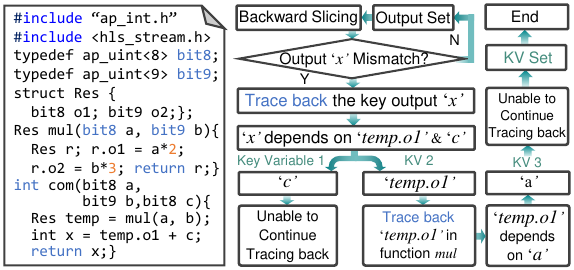}
	\vspace{-0.7cm}
	\caption{Using backward slicing to pinpoint key variables in the HLS program.}
	\label{fig:bs}
	\vspace{-0.25cm}
\end{figure}
\begin{figure}[t]
\centering
	\includegraphics[width=1\linewidth]{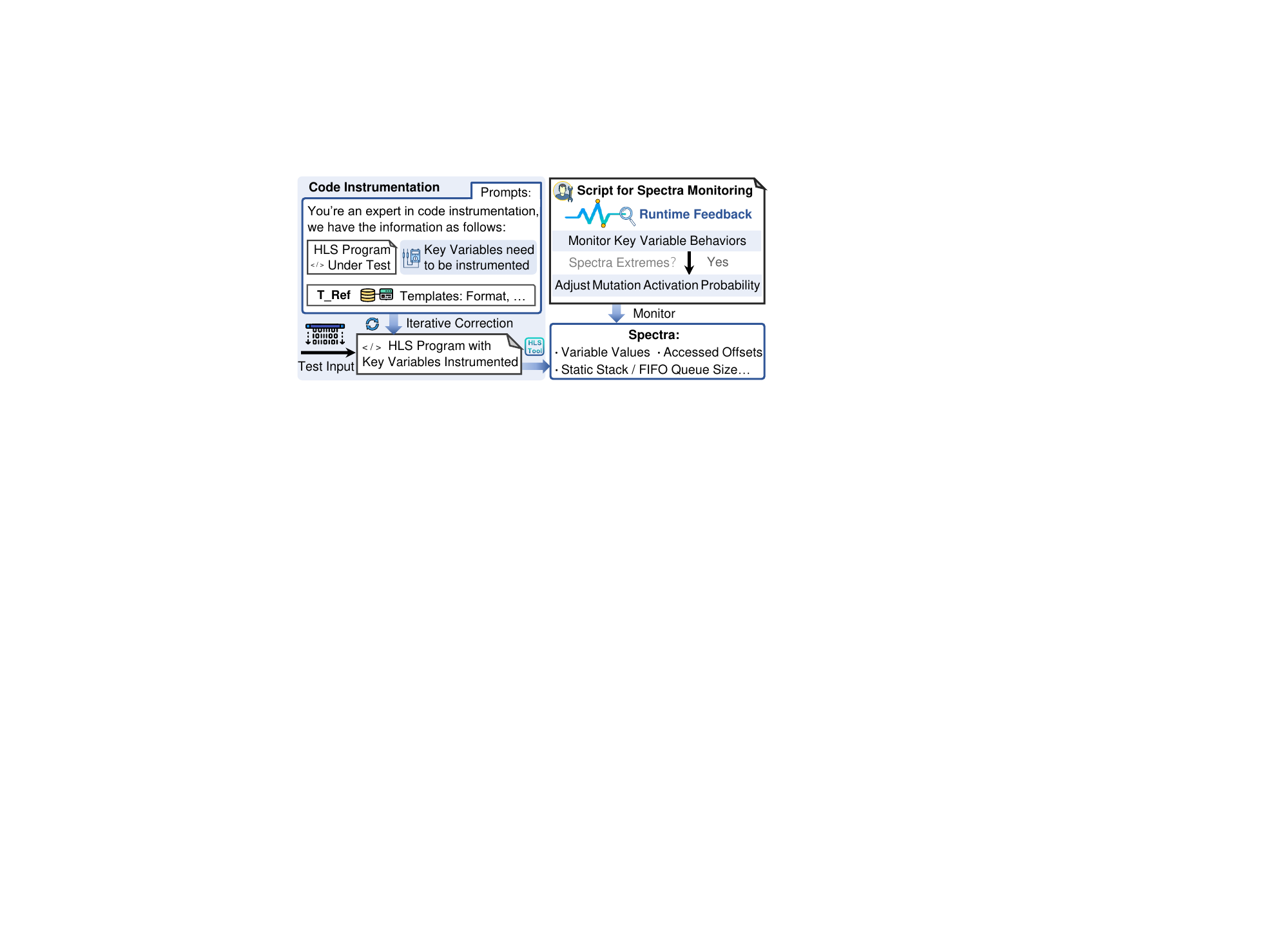}
	\vspace{-0.7cm}
	\caption{Code instrumentation and spectra monitoring for discrepancy analysis.}
	\label{fig:ci}
	\vspace{-0.6cm}
\end{figure}

\subsection{Instrumentation and Monitoring for Discrepancy Analysis}

To enable an in-depth analysis of discrepancies, a backward slicing technique, combined with the semantic understanding capabilities of LLMs, is employed to instrument key variables within the program, and monitoring scripts are developed to record their runtime spectra.
Backward slicing is a program analysis technique that traces data dependencies in reverse from a target output to identify all variables that affect its value \cite{b16.5}. For example, in a function that computes an output through a series of intermediate calculations, backward slicing can identify the relevant variables that contribute to the final result. Consider a dependency graph \(G = (V, E)\), where \(V\) is the set of variables and \(E\) is the set of dependency edges. Let \(x\) be the target output variable that needs to be traced back. The backward slicing process is then formalized as follows:
\vspace{-0.16cm}
\begin{equation}\label{eq:initialization}
S_{0} = \{\, x \}, S_{i+1} = S_{i}\cup\bigl\{\,u\in V\mid\exists\,v \in S_{i}:(u,v) \in E \bigr\},\vspace{-0.16cm}\end{equation}
where \(S_{0}\) starts with the target output variable \(x\), and with each iteration, any variable \(u\) that influences a variable \(v\) already in the current slice \(S_{i}\) is added to form the next slice \(S_{i+1}\). The slicing process is repeated until no variables can be added, i.e., when $S_{i+1} = S_{i}$, and the final slice is denoted by $S(x) = S_{i}$. 
%The set of key variables, denoted as $\mathrm{KV}$, is defined as all the variables in the backward slice.
% \vspace{-0.16cm}
% \begin{equation}
% \mathrm{KV}(x) = \{\, v \in S(x) \mid v \neq x \},
% \vspace{-0.16cm}
% \end{equation}
% meaning any variable that contributes (directly or indirectly) to the final output \(x\) is considered a key variable.

\begin{figure}[t]
\vspace{-0cm}
\centering
	\includegraphics[width=1\linewidth]{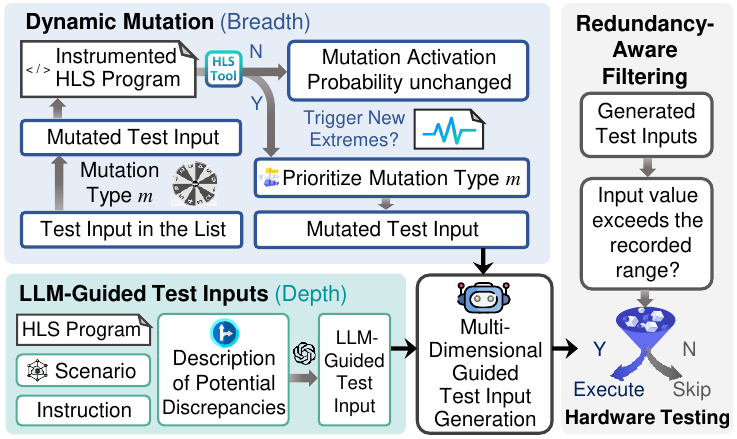}
	\vspace{-0.7cm}
\caption{Multi-dimensional test input generation and redundancy-aware filtering.}
	\label{fig:ti}
	\vspace{-0.65cm}
\end{figure}

To illustrate the process of the backward slicing technique, a simple example is presented in Fig.~\ref{fig:bs}. The process begins at the target output variable \textit{x}, where discrepancies between C/C++ execution and RTL simulation are observed. Analysis shows that \textit{x} depends on \textit{temp.o1} and \textit{c}. Since \textit{c} cannot be traced further, it is identified as a key variable (KV). The slicing then traces \textit{temp.o1} in the \textit{mul} function and reveals its dependence on \textit{a}, which is subsequently marked as a key variable. The final set of key variables is KV(x) = \{\textit{temp.o1, c, a}\}, and these variables are targeted for instrumentation and runtime monitoring. 

To facilitate runtime monitoring, our framework leverages an LLM to instrument the program using the previously pinpointed key variables. As shown in Fig.~\ref{fig:ci}, once the key variables are pinpointed, the HLS code is integrated with predefined instrumentation templates, which are provided to the LLM as prompts. These templates specify how to add instrumentation constructs and specify the output format for the runtime spectra. The LLM then generates an instrumented version of the HLS code that records the runtime spectra $Spectra(V)$ of these variables, which can be defined as follows:
$Spectra(V) = \{(\mathrm{val}(V),\, \mathrm{offset}(V),\, \mathrm{loop}(V),\, \mathrm{stack}(V),\, \mathrm{FIFO}(V))\}$, where $\mathrm{val}(V) = [\min(V),\,\max(V)]$ denotes the recorded value range, $\text{offset}(V)$ captures array offsets, $\text{loop}(V)$ records loop iteration counts, $\text{stack}(V)$ represents static stack usage, and $\text{FIFO}(V)$ tracks FIFO queue sizes. The runtime spectra are subsequently analyzed by a monitoring script, which generates detailed reports to highlight discrepancy symptoms such as overflow, division by zero, and truncation. Table~\ref{fig:type} summarizes the five categories of spectra currently supported by this framework; these mappings can be extended by modifying a configuration file. For instance, array access offsets are monitored to detect out-of-bounds behavior, and actual loop iteration counts are verified to ensure that loop unrolling optimizations align with expectations. Furthermore, the monitoring script reports a feedback array (e.g., \texttt{spectra\_feedback}) that records the type, name, minimum, and maximum value of each variable in Table \ref{fig:dm}. This feedback also provides a basis for subsequent dynamic mutation, which is detailed in the next subsection.

\vspace{-0.1cm}
\subsection{Multi-Dimensional Guided Test Input for Effective Testing} 
At this stage, two complementary techniques are integrated for test input generation: dynamic mutation and LLM-guided mechanism. As shown in Fig. \ref{fig:ti}, the dynamic mutation approach broadly explores the test input space by applying a diverse set of mutation operations, while the LLM-guided mechanism uses semantic analysis to generate test inputs that target program features. Through multiple rounds of validation, 30\% of test inputs are generated by the LLM, focusing on program insights, while the remaining test inputs are generated via dynamic mutation to ensure broad coverage. 

\subsubsection{Dynamic Mutation}
As different mutations vary in their ability to trigger different behaviors \cite{b16.6}, prioritizing effective mutation operations to test platform-dependent behaviors in HLS presents a significant challenge. In our framework, activation probabilities for mutation types are dynamically adjusted to prioritize those that produce new spectra values, thereby increasing the chances of revealing platform-dependent behavioral discrepancies. Since FPGA test inputs are typically arrays or matrices, our scheme employs eight mutation types to generate new test inputs:

\begin{table}[t]
  \vspace{0.15cm}
  \centering
  \refstepcounter{table}%
  {TABLE II: DYNAMIC MUTATION FOR TEST INPUT GENERATION.\par}
  \label{fig:dm}
  \vspace{0.1cm}
  \includegraphics[width=0.99\linewidth]{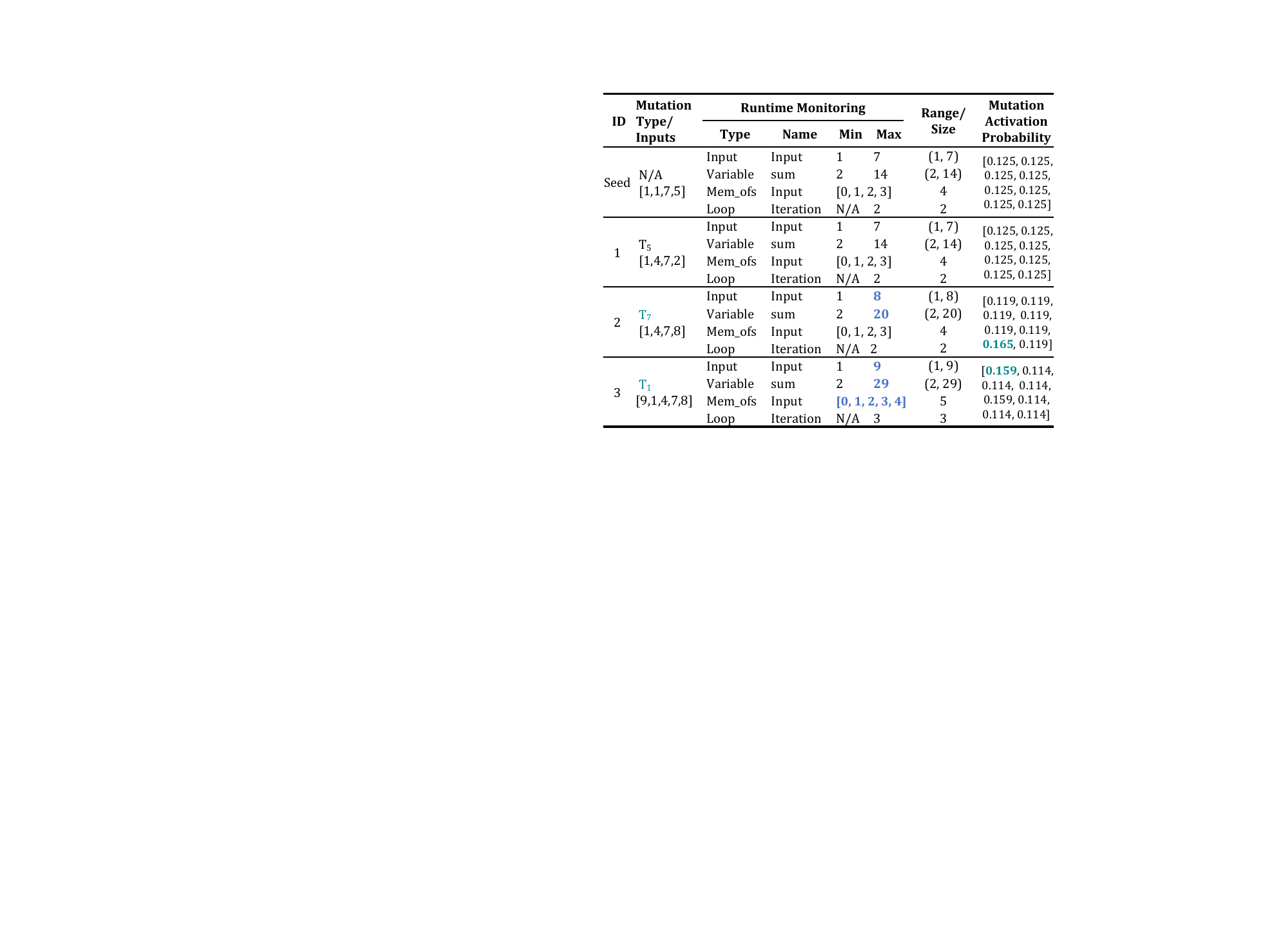}
  \vspace{-0.7cm}
\end{table}

\ding{172} \textit{Data Size Mutation}: Inserts or deletes elements in the test input, e.g., changing the input \texttt{[i, j]} to \texttt{[i, j, k]}.

\ding{173} \textit{Data Dimension Mutation}: Adjusts matrix dimensions, such as adding or removing a row/column, e.g., modifying \texttt{[[i, j], [m, n]]} to \texttt{[[i, j]]} or \texttt{[[i], [m]]}.

\ding{174} \textit{Zero Value Mutation}: Sets the elements to zero, e.g., changing an element of \texttt{[i, j]} to \texttt{0} to obtain \texttt{[0, j]}.

\ding{175} \textit{Order Mutation}: Rearranges the order of elements, e.g., transforming the array \texttt{[h, i, j]} to \texttt{[i, j, h]}.

\ding{176} \textit{Data Element Mutation}: Alters the values of elements, e.g., converting \texttt{[h, i, j]} to \texttt{[h, i, k]}.

\ding{177} \textit{Data Type Mutation}: Changes the data type while keeping the same value, e.g., converting integer \texttt{2} to floating-point \texttt{2.0}.

\ding{178} \textit{Bit Mutation}: Randomly flips a bit, e.g., \texttt{0101} to \texttt{0100}.

\ding{179} \textit{Byte Mutation}: Randomly flips a byte, e.g., turning the 16‑bit value \texttt{0x1A2B} into \texttt{0x1AD4} by flipping its lower byte.

Traditional fuzz testing typically assigns uniform activation probabilities to all mutation types, which restricts its adaptability to specialized hardware platforms. In contrast, our framework dynamically adjusts the activation probabilities in real-time according to monitored spectra feedback, prioritizing those mutations that trigger new spectra extremes.

Given $l=8$ types of mutation operations, each is initially assigned an activation probability of $P_m^{(0)} = \frac{1}{l} = 0.125$. In each iteration \textit{t}, a mutation operation 
$T_m$ is selected to generate an input. If the generated input triggers new extreme values (e.g., variable value ranges, array access offsets, or stack sizes, etc.), its activation probability increases as follows: 
\vspace{-0.13cm}
\begin{equation}
P_m^{(t+1)} = P_m^{(t)} + \alpha,
\label{eq:increase}
\vspace{-0.2cm}
\end{equation}
while the probabilities for other mutations decrease proportionally: 
\vspace{-0.32cm}
\begin{equation}
P_i^{(t+1)} = P_i^{(t)} - \frac{\alpha}{l-1}, \quad i \neq m,
\label{eq:decrease}
\vspace{-0.18cm}
\end{equation}
where $\alpha$ is a predefined update factor set at $0.04$. 
% These updates are subject to the following constraints:
% \vspace{-0.2cm}
% \begin{equation}
% \sideset{}{_{i=1}^{l}}\sum P_i^{(t+1)} = 1, 
% \quad 
% 0 \,\le\, P_i^{(t+1)} \,\le\, 1, 
% \quad \forall\,i \in \{1, 2, \dots, l\}.
% \label{eq:decrease}
% \vspace{-0.2cm}
% \end{equation}
For example, in the second iteration (transitioning from \textit{t}=1 to \textit{t}=2), if mutation operation $T_5$ generated a test input that expanded the spectra extreme value, its activation probability increases from $P_5^{(1)} = 0.125$ to $P_5^{(2)} = P_5^{(1)} + \alpha = 0.125 + 0.04 = 0.165$, while the other operations adjust to $P_i^{(2)} = 0.119 \quad (i \neq 3)$.

An example illustrating the dynamic mutation guided by spectra feedback is shown in Table~\ref{fig:dm}. The first column lists the test input IDs, while the second column displays the test inputs along with their corresponding mutation types $T_m$, where \( m \in [1, 8] \), and \( m \in \mathbb{Z}^+ \), representing the eight designed mutation types. Columns 3 to 6 record the monitored variable types, names, and the current extreme values captured by the spectra monitoring script, whereas column 7 presents the real-time spectra ranges of the variables. The final column indicates the activation probability assigned to each mutation. Initially, a test input seed [1, 1, 7, 5] is employed for HLS testing. In the first mutation round, each mutation type is assigned an activation probability of 0.125. Mutation type $T_5$ is activated, yielding a new test input (ID1: [1, 4, 7, 2]). Since the ranges of the variable spectra have not been expanded, the activation probabilities are maintained. Subsequently, mutation type $T_7$ is activated, producing a new test input (ID3: [1, 4, 7, 8]). At this point, the script detects that the variable \textit{sum} attains a new extreme value of 20, prompting an increase in the activation probability for mutation type $T_7$ in subsequent mutations. 

\subsubsection{LLM-Guided Test Input Generation}
\begin{figure}[t]
\vspace{-0.09cm}
\centering
	\includegraphics[width=0.97\linewidth]{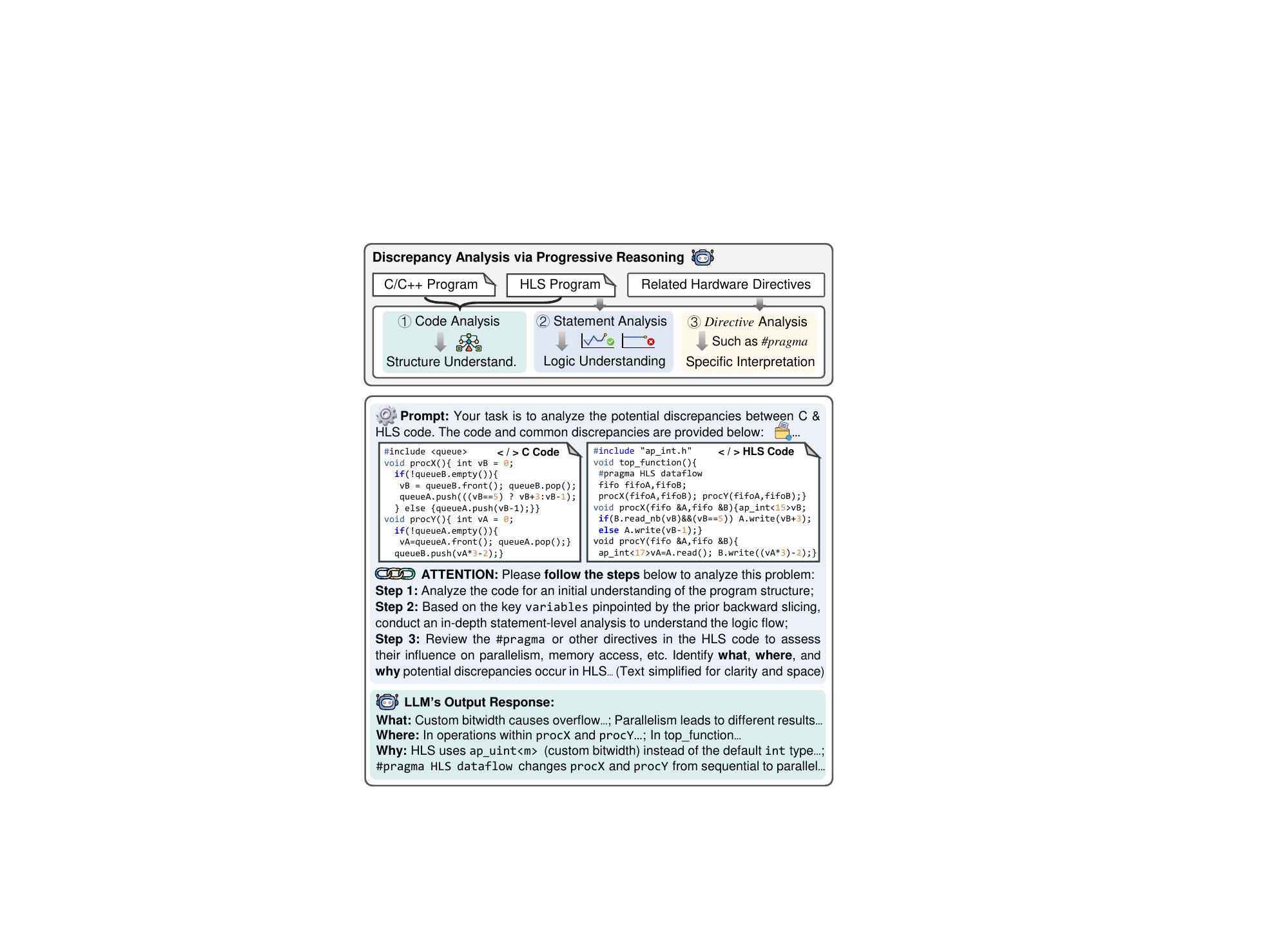}
	\vspace{-0.3cm}
	\caption{LLM-aided reasoning chain for progressive behavioral discrepancy analysis. Top: Execution steps. Bottom: Analysis example.}
	\label{fig:llmti}
	\vspace{-0.65cm}
\end{figure}

Although dynamic mutation effectively minimizes ineffective mutations while ensuring broad coverage of the test input space, relying solely on mutation may not sufficiently explore the program's details. 
At this stage, by leveraging the semantic understanding and adaptive learning capabilities of LLMs, an LLM-guided test input generation mechanism is further proposed. 
Inspired by the Chain-of-Thought technique, an LLM-driven reasoning chain is employed to progressively analyze discrepancies between an HLS program and its corresponding C/C++ program. As shown in Fig.~\ref{fig:llmti}, the process initiates with an overall code analysis to establish an understanding of the program structure. This is followed by a statement-level analysis that delves into the code logic by examining statements and function calls. Finally, a directive analysis phase focuses on hardware-specific interpretations, such as \texttt{\#typedef ap\_int<n>} and \texttt{\#pragma HLS pipeline}, that can customize bit width or alter the execution mode. An example of the LLM-driven reasoning chain, including the prompts and responses, is provided at the bottom of Fig.~\ref{fig:llmti}, illustrating how the framework analyzes potential discrepancies for the test input generation.
The considerations for employing LLMs to generate test inputs are as follows:

(1) HLS Program with Potential Discrepancies: Building on the reasoning chain, the LLM systematically dissects potential discrepancies uncovered during overall code analysis, individual statement analysis, and directive analysis and then strategically designs targeted test inputs that hone in on these critical discrepancy points.

(2) Task Scenario: The application scenario, such as data encryption, image processing, or matrix computations, is embedded in the LLM prompts, enabling the LLM to generate test inputs that are aligned with the testing objectives.

(3) Data Format Constraints: HLS programs often have strict data format requirements. The customized data type specifies both the variable type and its bit width \cite{b16.8} (e.g., \texttt{ap\_fixed$<$N$>$} on FPGA), while the data structure defines the format of the input data, such as a one-dimensional array or a two-dimensional matrix, along with its dimensions (e.g., \texttt{A[Row][Column]} on FPGA). The prompts provided to the LLM need to clearly specify these format constraints to ensure that the generated test inputs meet the required specifications. In addition, the prompt should explicitly describe how the test inputs should be arranged. For example, generating one test input per line, separating values with spaces, and placing them between triple backticks. This ensures that the output format is correct and prevents testing failures due to formatting issues.

\begin{table*}[t]
\vspace{-0.25cm}
\centering
  \refstepcounter{table}%
  {\MakeUppercase{TABLE III: Comparison of the proposed HLSTester with the GPT baseline and the traditional method.}\par}
	\vspace{0.15cm}
	\includegraphics[width=0.98\linewidth]{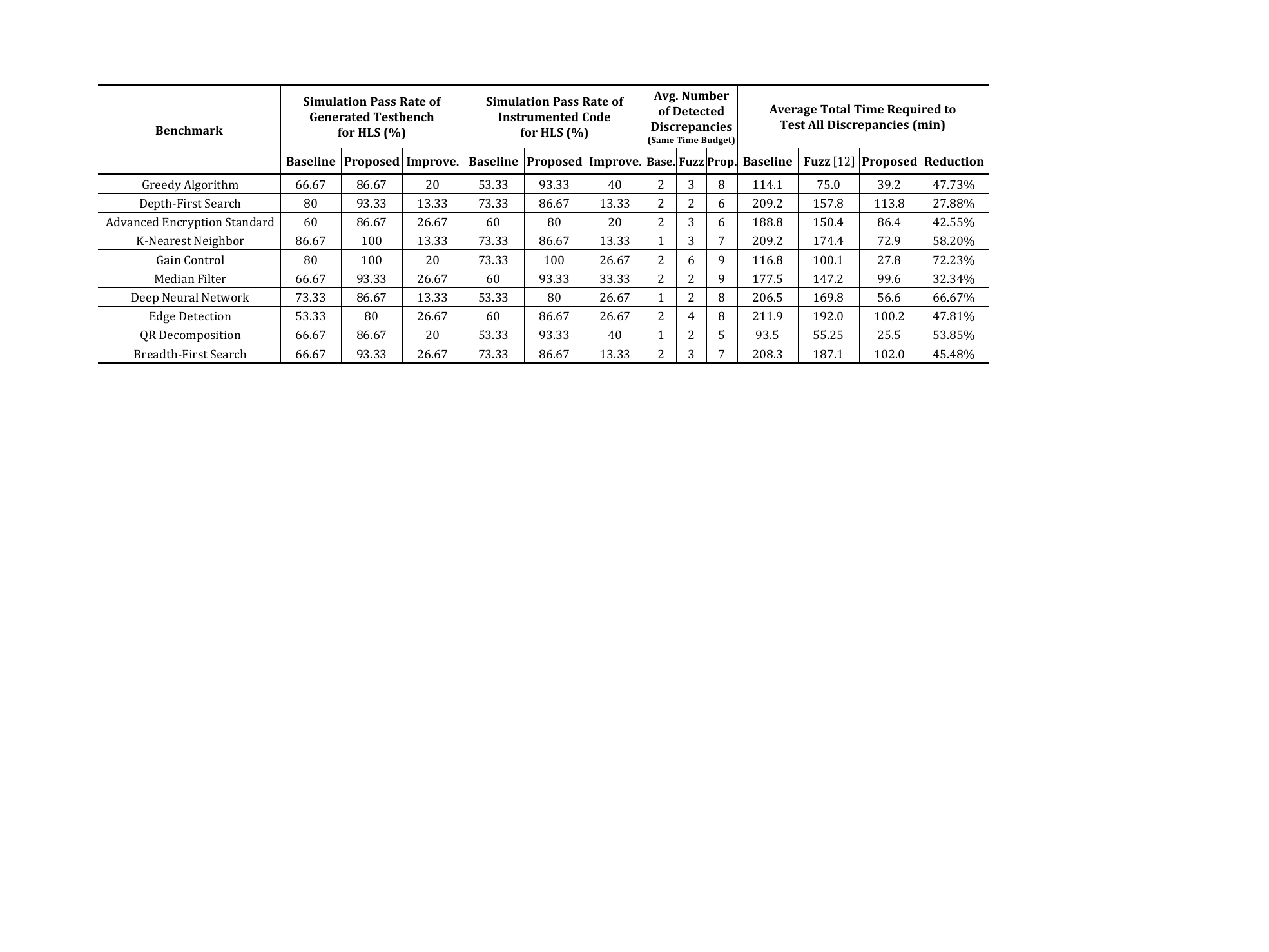}
	\label{tab:maint}
	\vspace{-0cm}
	\begin{tablenotes}
\item[] \scriptsize * The simulation pass rate and testing performance metrics are calculated from the results of 15 rounds of HLS simulation.
\end{tablenotes}
\vspace{-0.1cm}
\end{table*}

\begin{figure*}[t]
\vspace{-0.2cm}
\centering
	\includegraphics[width=0.985\linewidth]{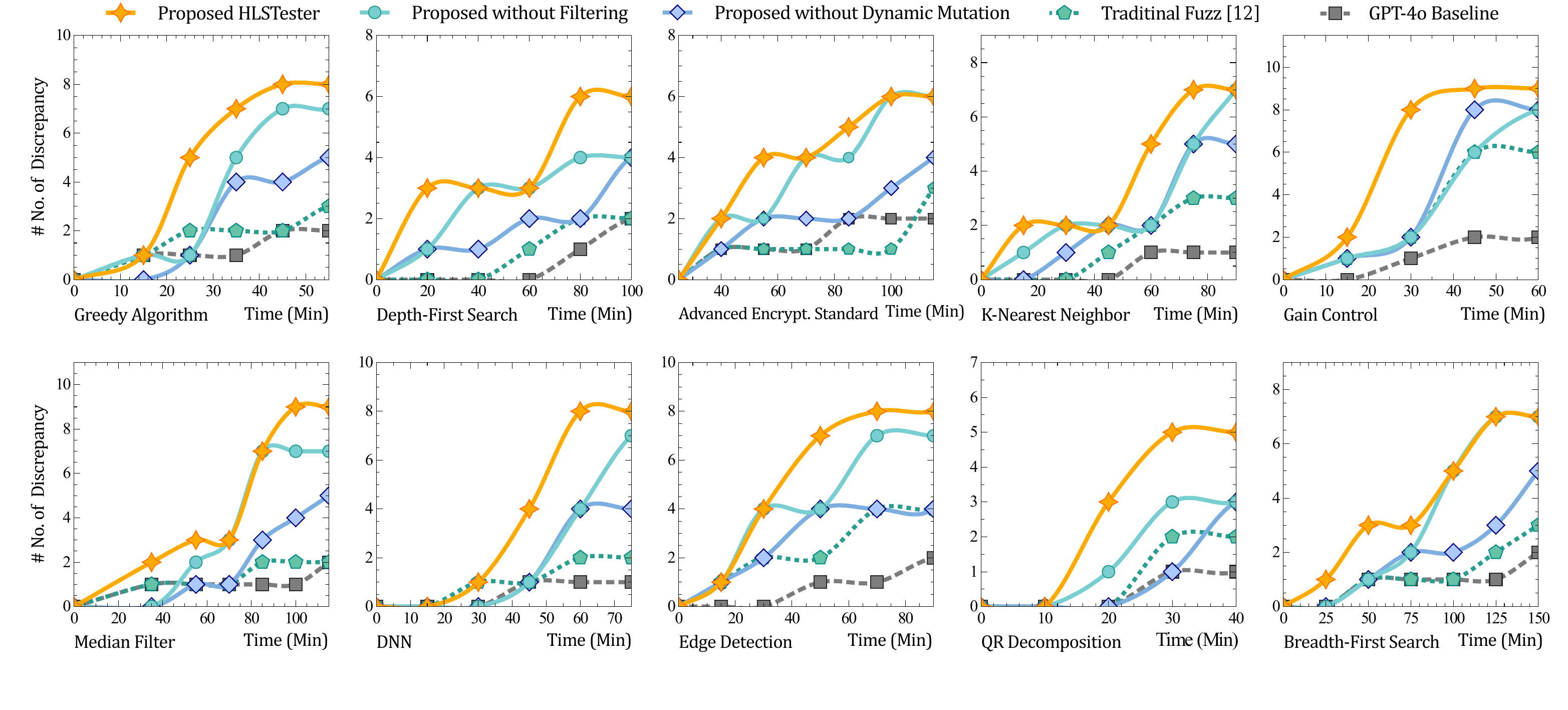}
	\vspace{-0.4cm}
	\caption{Comparison of the number of behavioral discrepancies detected within the same time budget across ten tasks, as shown on the x-axes of the figures.}
	\label{fig:input}
	\vspace{-0.65cm}
\end{figure*}

\vspace{-0.13cm}
\subsection{Redundancy Filtering of Test Inputs for Efficient Testing} 
\vspace{-0.08cm}
To avoid redundant simulations of test inputs that would exhibit the same behavioral discrepancies, a redundancy-aware technique is proposed to track the value ranges and data sizes of previously tested inputs. This technique dynamically maintains a global record table that stores the value ranges and data sizes of observed test inputs. When a new test input is generated, the framework first checks whether its value range or size exceeds those currently recorded. If the input's range or size extends beyond the recorded range, it is likely to trigger new behaviors and hardware simulation will be executed. If the test input remains within the previously recorded range, it implies that the test input will not trigger new discrepancies, and simulation is skipped. For example, assume that the current recorded input range is (2, 5). If the input is [1, 4, 5], where the minimum value is lower than the recorded range, simulation is executed, and the range is updated to (1, 5). However, if a subsequently generated input is [1, 3, 5], its range remains within (1, 5). Since inputs within the recorded range are likely to produce repetitive discrepancies, the hardware simulation could be skipped, thereby reducing unnecessary simulations.

\vspace{-0.1cm}
\section{Experimental Results}\label{sec:fourth}
We demonstrate the results of the proposed LLM-aided framework for HLS across 10 tasks in terms of the simulation pass rate of the modified testbench and instrumented code, the number of detected behavioral discrepancies within the same time budget, and the total time required for testing all behavioral discrepancies. Our benchmarks are from related work \cite{b16.1,b16.11,b16.2,b16.3} covering different hardware directives (e.g., customized bit width, memory management, parallelism, etc.).

\begin{figure}[t]
\vspace{-0cm}
\centering
	\includegraphics[width=1\linewidth]{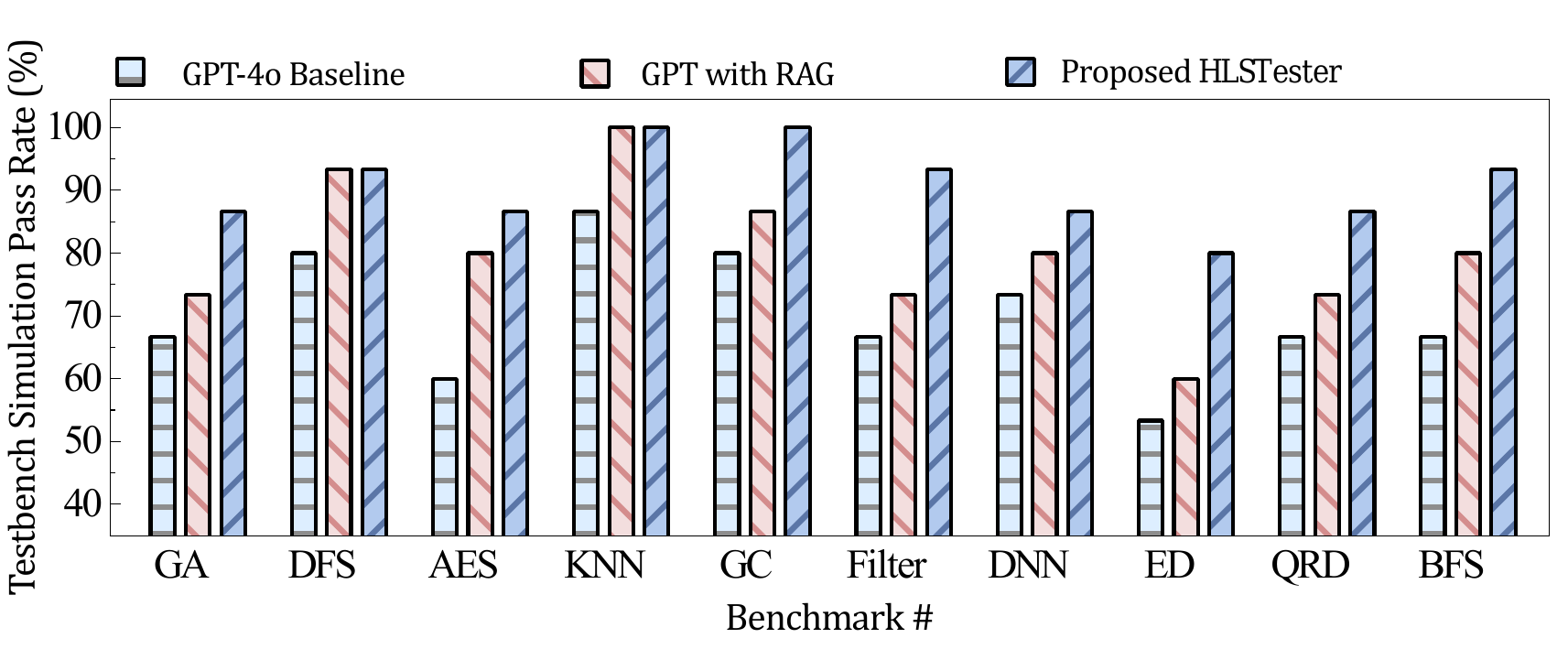}
	\vspace{-0.75cm}
	\caption{Comparison of the simulation pass rate of the modified HLS testbench among the proposed HLSTester, GPT baseline, and GPT with RAG.}
	\label{fig:tb}
	\vspace{-0.7cm}
\end{figure}

During the evaluation, because fine-tuning LLMs is not cost-effective in many design scenarios, the GPT-4o model was employed as the LLM via OpenAI APIs \cite{b16.4}, as it represents a typical and generalized LLM for broad tasks without requiring fine-tuning. Moreover, using general LLMs such as GPT-4o allows the HLS process to automatically benefit from continual improvements as the models are updated. All experiments were conducted using the AMD Vitis HLS Tool to simulate program execution on a Xilinx Virtex UltraScale+ XCVU9P-FLGA2104I FPGA with an Intel Xeon Silver 4314 CPU. To ensure the reliability and statistical robustness of our evaluation, each experiment for a specific task was repeated $n$ instances ($n$ = 15). In each instance, the LLM was queried five times to iteratively correct the program based on the feedback from the HLS tool. We calculate the pass rate as $\text{Pass Rate (\%)}= m/n$, where $m$ represents the number of successfully generated instances and $n$ denotes the total number of instances.
\begin{figure}[t]
\vspace{-0.5cm}
\centering
\includegraphics[width=0.99\linewidth]{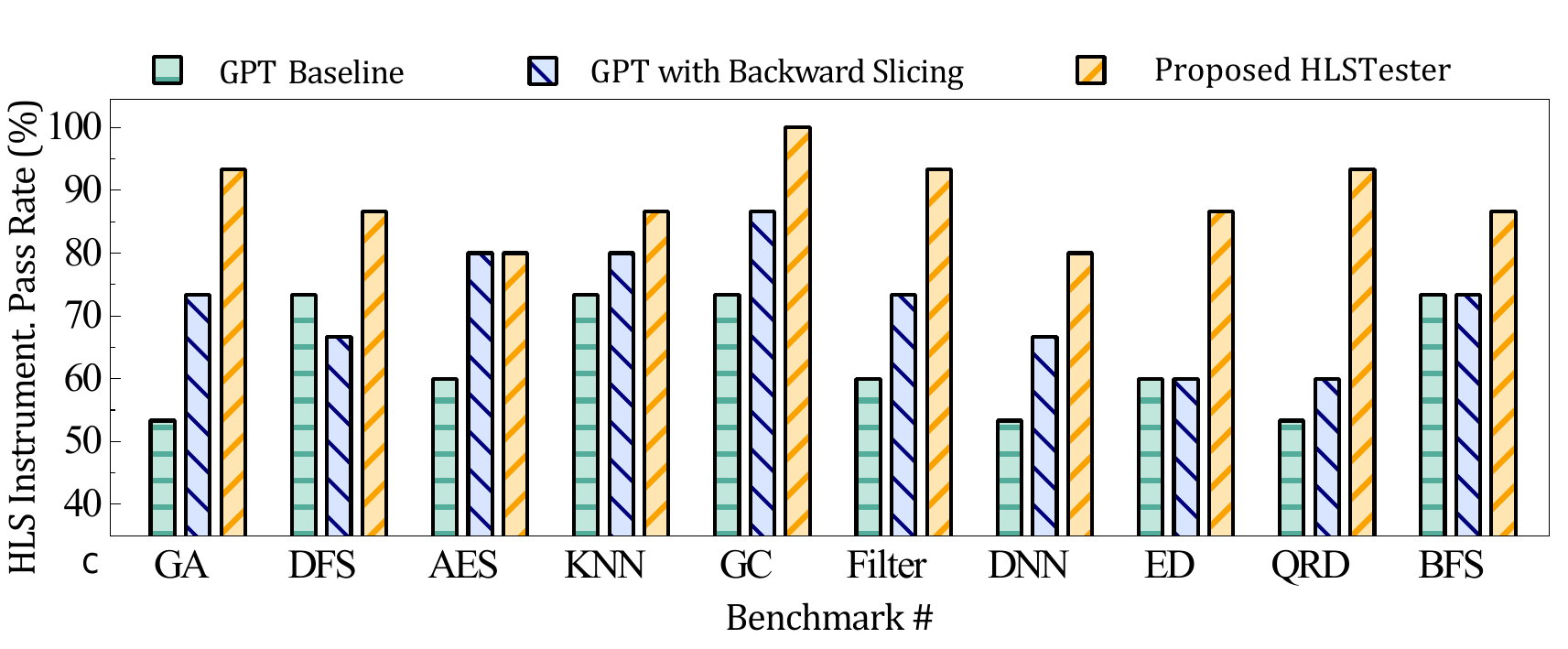}
\vspace{-0.75cm}
\caption{Comparison of the simulation pass rate of the instrumented HLS code among the proposed method, GPT baseline, and GPT with backward slicing.}
\label{fig:ih}
\vspace{-0.35cm}
\end{figure}

\begin{figure}[t]
%\vspace{-0.1cm}
\centering
\includegraphics[width=0.99\linewidth]{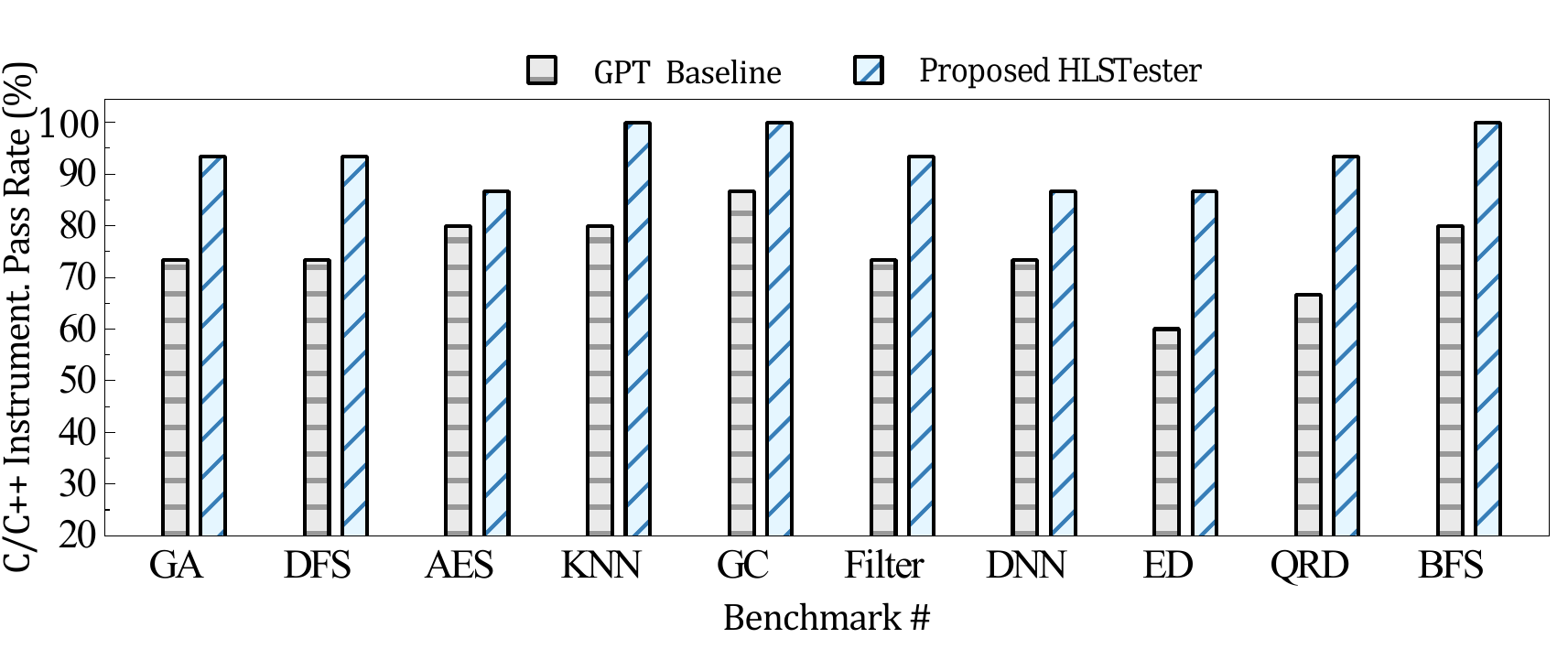}
\vspace{-0.75cm}
\caption{Comparison of the pass rate for instrumented C++ code between the proposed method and the GPT baseline.}
\label{fig:ic}
\vspace{-0.75cm}
\end{figure}

Table~\ref{tab:maint} compares the performance of the proposed HLSTester with the GPT baseline and the traditional testing method~\cite{b8.1}. The first column lists the name of benchmark tasks, while the second through seventh columns show the simulation pass rates of the modified testbench and instrumented code for HLS, with the proposed framework demonstrating superior performance compared to the baseline. The eighth through tenth columns report the average number of detected behavioral discrepancies, demonstrating that the proposed HLSTester identifies more discrepancies within the same time budget. The final four columns detail the total time required to test all discrepancies and the corresponding reductions. According to these columns, the proposed HLSTester outperforms both the GPT baseline and the traditional method by achieving a higher pass rate and shorter overall testing time.

To demonstrate the effectiveness of the proposed method in testbench modification, we compare the simulation pass rate of the proposed HLSTester with using GPT directly. According to Fig. \ref{fig:tb}, the HLSTester outperforms the baseline by achieving an average 20.67\% increase in testbench modification.

Fig. \ref{fig:ih} and Fig. \ref{fig:ic} present comparisons of the simulation pass rate for code instrumentation, where the proposed framework was compared with the direct use of GPT and GPT with backward slicing. In these comparisons, the proposed HLSTester demonstrated average improvements of 25.33\% and 18.67\% in HLS and C/C++ code instrumentation, respectively. 

To demonstrate the acceleration achieved by the HLSTester in generating effective test inputs for efficient discrepancy detection, four baselines including ablation options were created by downgrading HLSTester in Fig. \ref{fig:input}, summarized as follows:

$\circ$~\textit{GPT Baseline:} This baseline employs GPT directly to generate test inputs to detect behavioral discrepancies.

$\circ$~\textit{Traditional Fuzz Testing:} In this option, the HLS tool is invoked for every test input generated by fuzz testing.

$\circ$~\textit{Proposed without Mutation:} This option disables the proposed mutations from HLSTester. %to demonstrate the role of dynamic mutations in increasing the chance of revealing behavioral discrepancies.

$\circ$~\textit{Proposed without Redundancy-Aware Filtering:} In this option, redundancy filtering is disabled in HLSTester.

\begin{figure}[t]
\vspace{-0.5cm}
\centering
	\includegraphics[width=0.99\linewidth]{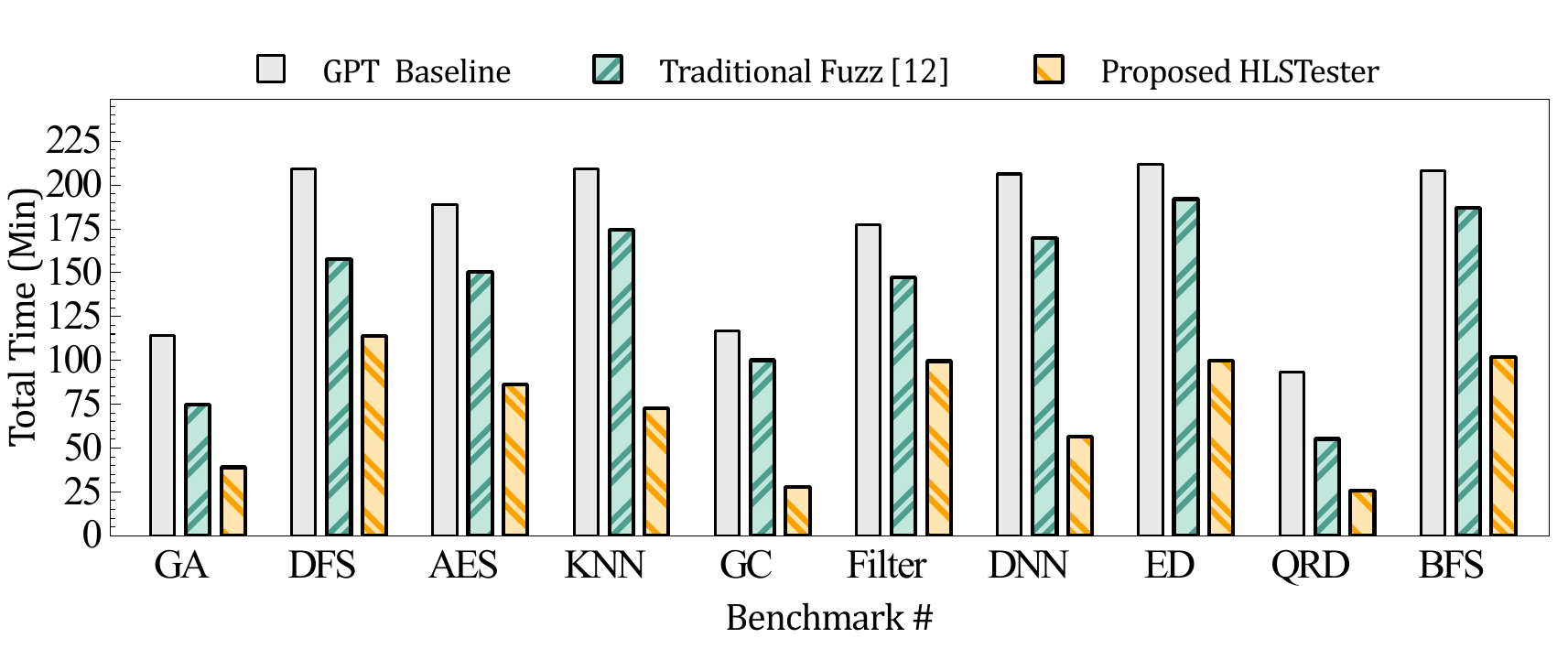}
	\vspace{-0.75cm}
	\caption{Comparison of the average total time required to test all discrepancies.}
	\label{fig:time}
	\vspace{-0.7cm}
\end{figure}

Fig.~\ref{fig:input} compares the proposed HLSTester framework with four baselines in terms of behavioral discrepancy detection speed. The x-axis denotes the elapsed time (i.e., the testing time budget), while the y-axis shows the average cumulative number of detected discrepancies. By incorporating an LLM-driven reasoning chain, dynamic mutation, and redundancy filtering, the proposed HLSTester generates more effective test inputs to detect more behavioral discrepancies in the same time budget. In particular, the LLM-driven reasoning chain achieves an average of 1.81× speedup in detecting all behavioral discrepancies, while the dynamic mutation increases the speedup to an average of 2.28x, and the redundancy‑aware technique further accelerates the testing workflow by 15.73\%.

To evaluate the total testing time required to detect all behavioral discrepancies, Fig. \ref{fig:time} compares the proposed HLSTester with the direct use of GPT and traditional fuzz testing. Each bar represents the average total testing time per benchmark, which includes HLS tool synthesis and simulation time, GPT interaction time, script execution time, and so on. As illustrated, the proposed HLSTester achieves an average speedup of 2.71$\times$ over the baseline, demonstrating the high efficiency achieved by optimizing the testing workflow.

\vspace{-0.1cm}
\section{Conclusion}\label{sec:fifth}
%In this paper, we propose HLSTester, an LLM-aided testing framework designed to efficiently detect behavioral discrepancies in HLS. Existing C/C++ testbenches are used to guide the LLM in modifying them into HLS-compatible versions. Then, a backward slicing technique is employed to pinpoint the key variables, and a monitoring script is developed to record their runtime spectra, thereby enabling a detailed analysis of discrepancy symptoms. Moreover, a test input generation mechanism is proposed that leverages spectra feedback to guide dynamic mutation, which is incorporated with insights from an LLM-based progressive reasoning chain to increase the likelihood of revealing behavioral discrepancies. In addition, a redundancy-aware technique is employed to skip repetitive simulations. Experimental results show that the proposed HLSTester significantly accelerates the testing workflow and achieves higher simulation pass rates compared with both the traditional method and the direct use of the LLM.

In this paper, we propose HLSTester, an LLM-aided testing framework designed to efficiently detect behavioral discrepancies in HLS. 
Existing C/C++ testbenches are used to guide the LLM in modifying them into HLS-compatible versions, with efficient test inputs generated by dynamic mutation with an LLM-based progressive reasoning chain to increase the likelihood of revealing behavioral discrepancies. 
A backward slicing technique is employed to pinpoint the key variables, and a monitoring script is developed to record their runtime spectra for detailed discrepancy analysis.%, thereby enabling a detailed analysis of discrepancy symptoms. 
Experimental results show that the proposed HLSTester significantly accelerates the testing workflow and achieves high simulation pass rates.

\vspace{-0.1cm}
\section*{Acknowledgement}
This work is funded by the Deutsche Forschungsgemeinschaft \textit{(DFG, German Research Foundation)} – Project-ID 504518248 and supported by TUM International Graduate School of Science and Engineering \textit{(IGSSE)}.

\end{document}